\documentclass[conference]{IEEEtran}
\pdfoutput=1
\usepackage[utf8]{inputenc}
\usepackage[T1]{fontenc}
\usepackage{amsmath,amssymb}
\usepackage{hyperref}
\usepackage{amsmath}
\usepackage{amsfonts}
\usepackage{tikz}
\usepackage{lipsum}
\usetikzlibrary{quantikz2}
\usepackage{algpseudocode,algorithm}
\begin{document}

% Title
% \title{Simulating Distributed Quantum Computing with Superconducting Qubits: A Performance Analysis of Noisy Optical Quantum Links vs. Classical Links}

\title{ Gate Teleportation vs Circuit Cutting in Distributed Quantum Computing}

% Authors

\author{\IEEEauthorblockN{Shobhit Gupta\IEEEauthorrefmark{2},
Nikolay Sheshko\IEEEauthorrefmark{2},
Daniel J. Dilley\IEEEauthorrefmark{3}, 
Alvin Gonzales\IEEEauthorrefmark{3},
Manish K. Singh\IEEEauthorrefmark{2}, and
Zain H. Saleem\IEEEauthorrefmark{3}
}
\IEEEauthorblockA{\IEEEauthorrefmark{2} memQ Inc., Chicago, IL 60615, USA}
\IEEEauthorblockA{\IEEEauthorrefmark{3}Math and Computer Science Division, Argonne National Laboratory, Lemont, IL 60439 USA}

% \thanks{Manuscript received December 1, 2012; revised August 26, 2015. 
% Corresponding author: M. Shell (email: http://www.michaelshell.org/contact.html).}
}

% \author{Shobhit Gupta}
% \affiliation{memQ, Inc, Chicago, IL 60615, USA}
% \author{Nikolay Sheshko}
% \affiliation{memQ, Inc, Chicago, IL 60615, USA}
% \author{Daniel J. Dilley}
% \affiliation{Math and Computer Science Division, Argonne National Laboratory, Lemont, IL 60439 USA}
% \author{Alvin Gonzales}
% \affiliation{Math and Computer Science Division, Argonne National Laboratory, Lemont, IL 60439 USA}
% \author{Manish K. Singh}
% \affiliation{memQ, Inc, Chicago, IL 60615, USA}
% \author{Zain H. Saleem}
% \affiliation{Math and Computer Science Division, Argonne National Laboratory, Lemont, IL 60439 USA}

\maketitle
% \thispagestyle{plain}
% \pagestyle{plain}
% Abstract
%
\begin{abstract}
    \noindent Distributing circuits across quantum processor modules will enable the execution of circuits larger than the qubit count limitations of monolithic processors. While distributed quantum computation has primarily utilized circuit cutting, it incurs an exponential growth of sub-circuit sampling and classical post-processing overhead with an increasing number of cuts. The entanglement-based gate teleportation approach does not inherently incur exponential sampling overhead, provided that quantum interconnects of requisite performance are available for generating high-fidelity Bell pairs. Recent advances in photonic entanglement of atomic qubits have motivated discussion on optical link metrics required to achieve remote gate performance approaching circuit-cutting techniques. We model noisy remote (teleported) gates between superconducting qubits entangled via noisy microwave-to-optical (M2O) transducers over optical links. We incorporate the effect of the transducer noise added ($N_{add}$) on the Bell pair state preparation circuit and inject noisy Bell pairs into remote CNOT gates. We perform comparative simulation of Greenberger–Horne–Zeilinger (GHZ) states generated between processor modules using remote gates and gate cuts by studying the dependence of the Hellinger fidelity on the primary source of error for the two approaches. We identify break-even points and regimes where noisy remote gates achieve parity with gate-cuts. In particular, our work suggests that a 10-fold reduction in the present M2O transducer noise added figures would favor generating multipartite entangled states with remote gates over circuit cutting due to an exponential sampling overhead for the latter. Our work informs near-term quantum interconnect hardware metrics and motivates a network-aware hybrid quantum-classical distributed computation approach, where both quantum links and circuit cuts are employed to minimize quantum runtime. 
    
\medskip

\end{abstract}

\small

% Introduction
\section{Introduction}
Near-term quantum computers have limited qubit counts and are prone to errors, which restrict their application in executing deep circuits with a large number of qubits. Expanding modules to larger sizes can allow increasing the number of qubits inside the module; however, this monolithic architecture has its challenges across qubit modalities. This includes heating and spectral crowding of motional modes in the case of trapped ions \cite{PhysRevX.14.041017}; cryostat size \cite{Krinner2019-bq}, qubit crosstalk, fabrication yield, and signal routing for superconducting qubits \cite{PRXQuantum.5.030350,mohseni2025buildquantumsupercomputerscaling}; and limited laser power and field of view constraints for neutral atoms \cite{Manetsch2025-ik,PhysRevX.13.041034}. Distributed quantum computing enables scaling beyond these limitations of a monolithic processor by interconnecting modules over optical quantum links \cite{BARRAL2025100747}. 

 Optical links have been proposed for networking modular quantum processors across several modalities \cite{PhysRevX.4.041041,chung2025interqnetheterogeneousfullstackapproach}, including neutral atom \cite{PRXQuantum.5.020363,PhysRevResearch.7.013313,PRXQuantum.6.010101}, trapped ion \cite{PhysRevA.89.022317}, solid-state \cite{PRXQuantum.5.010102}, and superconducting qubits \cite{Ang2024-fh}; where qubits across modules are entangled through photon-mediated heralded Bell state measurement. Photonic entanglement of neutral atoms and trapped ions has been demonstrated with fidelity of 94 -97 \% \cite{PhysRevLett.133.090802,Saha2025,PhysRevLett.124.110501}, and link rates up to $\sim$ 250 Hz \cite{PhysRevLett.133.090802}, along with a recent demonstration of gate teleportation over optical links \cite{Main2025}. While atom/ion-based qubits naturally interface with optical photons, microwave-domain qubits such as superconducting qubits require milliKelvin cryogenic links \cite{Almanakly2025-lz,Chou2018-ka,Yam2025-kk}. Due to minimal photon loss and thermal noise in optical fibers at room temperatures, optical links via microwave-to-optical (M2O) transducers are a suitable approach for enabling connectivity between qubits across dilution refrigerators \cite{Ang2024-fh,Mirhosseini2020-kx,Bravyi2022-zi}. Optical entanglement of superconducting qubits has yet to be demonstrated; however, there have been significant improvements in M2O transducer properties across several metrics \cite{weaver2025scalablequantumcomputingoptical}. State-of-the-art transducers are expected to produce probabilistic, noisy Bell pairs with fidelity of $\sim$50 \% \cite{dirnegger2025montecarlomodeldistilled}.

At present, optical links of sufficient rates and fidelity do not exist; therefore, quantum circuits have been partitioned over modules using classical circuit cutting techniques \cite{Tang_2021,PhysRevLett.130.110601}. Circuit cutting enables the execution of circuits larger than the QPU size by dividing them into sub-circuits, which are then individually sampled to recreate the quasiprobability distribution of a gate or a wire-cut. We refer to circuit-cutting as ``classical links" in this work, where the data from each module is post-processed asynchronously. Cutting-based approaches incur higher overhead in terms of quantum runtime from the sampling of sub-circuits and exponential classical post-processing overhead due to knitting of the cuts.   

While circuit cutting techniques have been employed for running algorithms on noisy intermediate scale quantum (NISQ) era processors, building large-scale, fault-tolerant modular quantum processors will require high-rate, high-fidelity optical quantum links \cite{PhysRevResearch.7.013313,sakuma2024opticalinterconnectmodularquantum,PRXQuantum.5.020363}. Current photonic entanglement rates for atomic qubits are slow ($\sim$ 250 Hz), thus motivating extensive research on distributed circuit compiler design to minimize the Bell pair budget, through efficient algorithm partitioning, remote-gate scheduling, and gate reordering \cite{Cuomo_2023,Wu2023entanglement,FerrariBandiniAmoretti,mengoni2025efficientgatereorderingdistributed,autocomm10.1109/MICRO56248.2022.00074,Ferrari_2023,Wu2023-iy}. Other works have studied distributed quantum error correction codes and schemes  \cite{singh2024modulararchitecturesentanglementschemes,sutcliffe2025distributedquantumerrorcorrection} as well as error-correcting thresholds for noisy quantum links \cite{Ramette2024-jf}. Entanglement-mediated remote gates have been simulated to enable long-range interaction within a processor module, facilitating optimum qubit routing \cite{padda2024improvingqubitroutingusing}. Works on distributed computing with remote gates so far have simulated Bell pair noise through a generic depolarization channel \cite{Muralidharan2025}. A recent study has emulated noisy optical interconnects on IBM quantum processors by modeling noise as an amplitude damping channel \cite{elyasi2025frameworkquantumdatacenter}. To establish the metrics for optical interconnects, it is critical to translate the physical hardware properties to remote gate errors and finally circuit-level errors. The noise sources in the optical interconnect can include noise photons added by the optical channel or by the M2O transducer, polarization mixing for photons emitted by atomic qubits, as well as local sources of qubit gate errors during Bell pair generation. While the long-term goal is to achieve remote (optical) link fidelity matching local (e.g., microwave) links \cite{dirnegger2025montecarlomodeldistilled}, a near-term metric would be achieving parity with circuit-cutting techniques \cite{Ang2024-fh}.

 In this work, we perform a comparative simulation of GHZ states generated between superconducting qubits across modules via remote (teleported) CNOT gates over quantum links and CNOT gate-cuts. The quantum link is generated by heralding Bell pairs between superconducting qubits interconnected via noisy quantum transducers over optical fibers. We then compare the Hellinger fidelity between remote gates and gate-cuts as a function of the GHZ circuit size (number of CNOT gates). We find that with transducer noise added ($\mathrm{N_{add}}$) of 0.01-0.1, remote gates can achieve GHZ fidelity comparable with the circuit-cutting approach for a fixed shot budget, with multipartite entangled state generation being advantageous for remote gates. This noise figure is within 10-fold of  state-of-the-art ($\mathrm{N_{add}} \sim$ 0.12-0.14 \cite{weaver2025scalablequantumcomputingoptical,Meesala2024-mk}). Our work addresses the near-term optical interconnect hardware metrics required to achieve utility in a hybrid quantum-classical approach \cite{mohseni2025buildquantumsupercomputerscaling}, where sparse quantum links can be used in conjunction with circuit-cutting techniques.

\section{Methods}

Our work models the dominant source of error for gate teleportation and circuit-cutting approaches and performs a comparative analysis by simulating the circuit fidelity as a function of the error parameter. The primary source of error for remote gates is the Bell pair infidelity from noisy quantum links. In the case of circuit cutting techniques, for a fixed shot budget (number of repetitions for each sub-circuit), the fidelity degrades with increasing number of cuts. This is because maintaining target fidelity requires an exponentially larger number of samples for each sub-circuit. We anticipate a regime where remote gates can offer comparable fidelity to gate-cuts, especially for generating inter-module multipartite entangled states, which require multiple non-local CNOT gates and thus incur larger sampling overhead. We do not constrain additional resource overheads in our analysis, including the time latency associated with entanglement generation for remote gates and classical post-processing overhead associated with circuit-cutting. Due to the different classical and quantum resource overheads between the two approaches, circuit fidelity serves as a suitable initial metric for comparison. Among the additional error sources, we implement local gate errors as a depolarization channel with 1-qubit gate fidelity of 0.99 and 2-qubit gate fidelity of 0.98. 

We simulate GHZ states generated between qubits across modules using non-local CNOT gates. GHZ states serve as primitives of multipartite entangled states \cite{Li:19}, which can be utilized for more sophisticated algorithms distributed across processor modules. The fidelity of GHZ states serves as a benchmark for the performance of intermodule CNOT gates, especially in the context of the degradation of GHZ state fidelity with a larger number of qubits, i.e, more non-local CNOT gates. GHZ states have been used to benchmark multipartite entanglement across superconducting qubit modules connected via cryogenic microwave links, with reported four-qubit intermodule GHZ state fidelity of 92 \% \cite{Niu2023-um}. 

\begin{figure}[!htpb]
    \centering
    \includegraphics[width=0.99\linewidth]{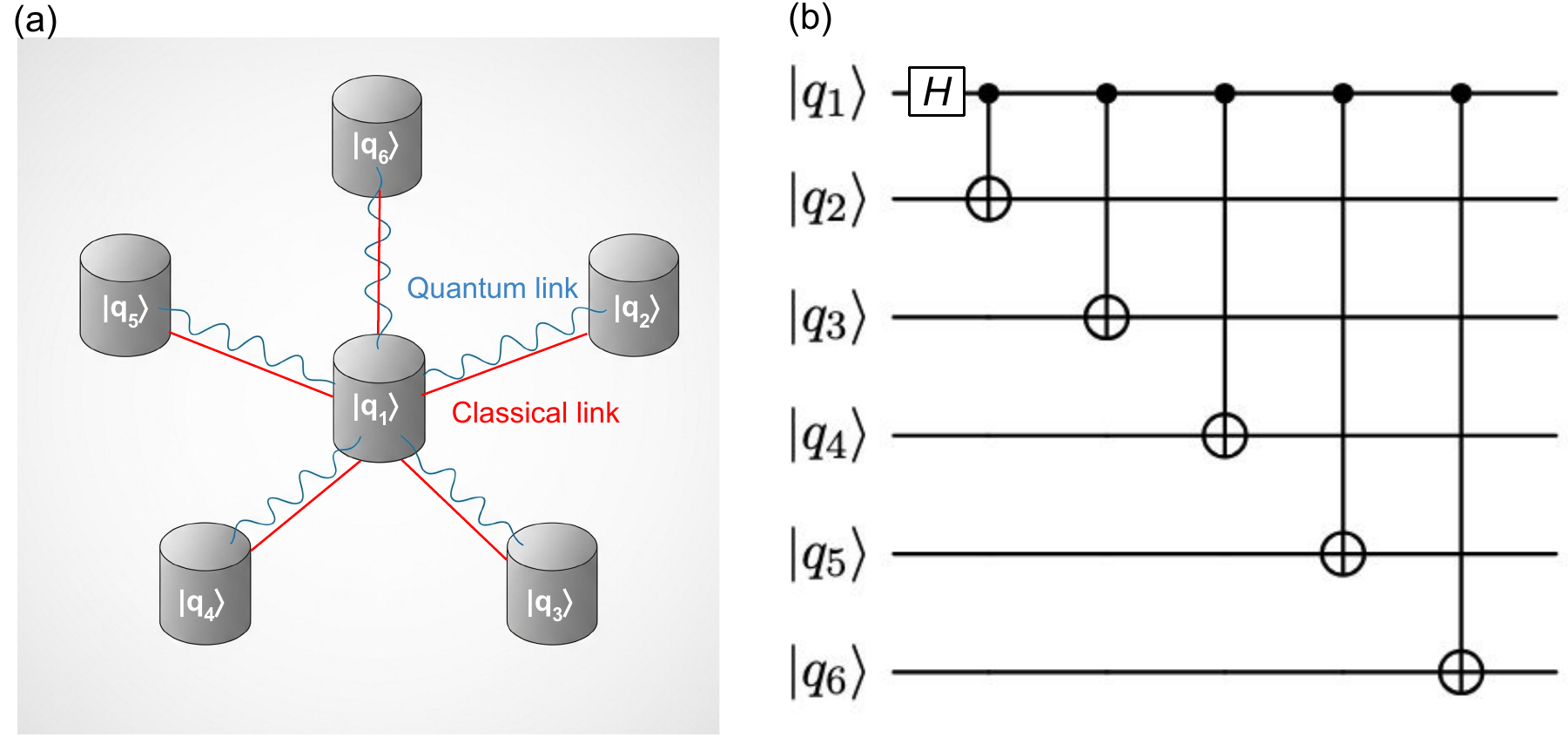}
    \caption{(a) Distributed processor architecture, quantum links are established through photonic Bell state measurements over optical fiber, while classical links are established through circuit-cutting techniques. (b) Intermodule GHZ N circuit generated using non-local CNOT gates using remote gates or gate-cuts.}
    \label{fig:1}
\end{figure}

Fig.~\ref{fig:1} shows the distributed processor architecture used in this work: a central module connected with five surrounding modules, each housing one data qubit. The modules are connected via entanglement-based optical links (quantum links, blue) and classical links (red). The QPUs are assumed to be in a datacenter scale setting where optical link distance is $\sim$10 m; therefore, the optical channel loss and latency are negligible. The non-local CNOT between the qubits is implemented via gate-cuts for classical links and remote CNOT gates for quantum links. 

% \begin{figure}[!htpb]
%     \centering
%     \begin{quantikz}
%         \lstick{\ket{q_1}} & \ctrl{1} & \ctrl{2} & \ctrl{3} & \ctrl{4} & \ctrl{5} & \\
%         \lstick{\ket{q_2}} & \targ{1} & & & & & \\
%         \lstick{\ket{q_3}} &  &  \targ{1} & & & &  \\
%         \lstick{\ket{q_4}} &  &  & \targ{1} & &  & \\
%         \lstick{\ket{q_5}} &  &  & & \targ{1} & &   \\
%         \lstick{\ket{q_6}} &  &  & &  &\targ{1} &  \\
%     \end{quantikz}
%     \caption{GHZ 6 circuits used in the simulation}
%     \label{fig:tcanther}
% \end{figure}

\subsection{Remote gate}

\begin{figure}[!htpb]
    \centering
    \includegraphics[width=0.99\linewidth]{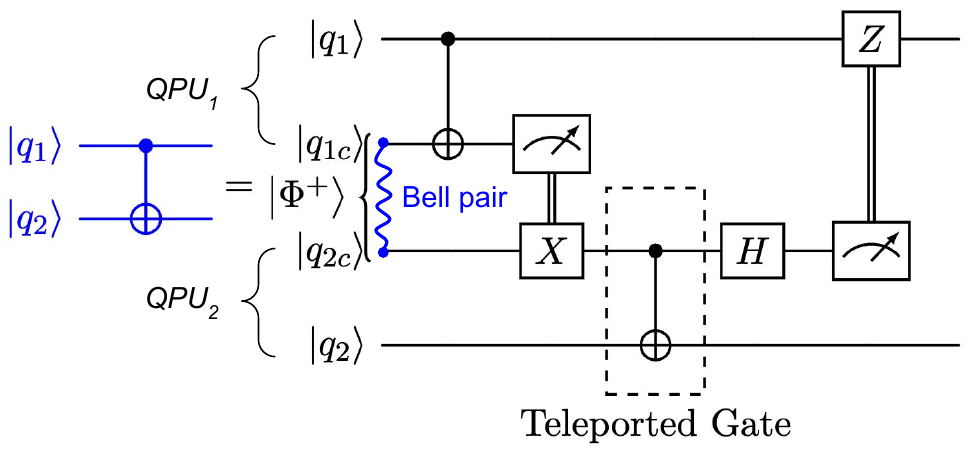}
    \caption{Remote CNOT telegate between qubits $\ket{q_1}$ and $\ket{q_2}$ implemented via Bell pairs shared between local communication qubits $\ket{q_{1c}}$ and $\ket{q_{2c}}$.}
    \label{fig:2}
\end{figure}

\begin{figure}[!htpb]
    \centering
    \includegraphics[width=0.99\linewidth]{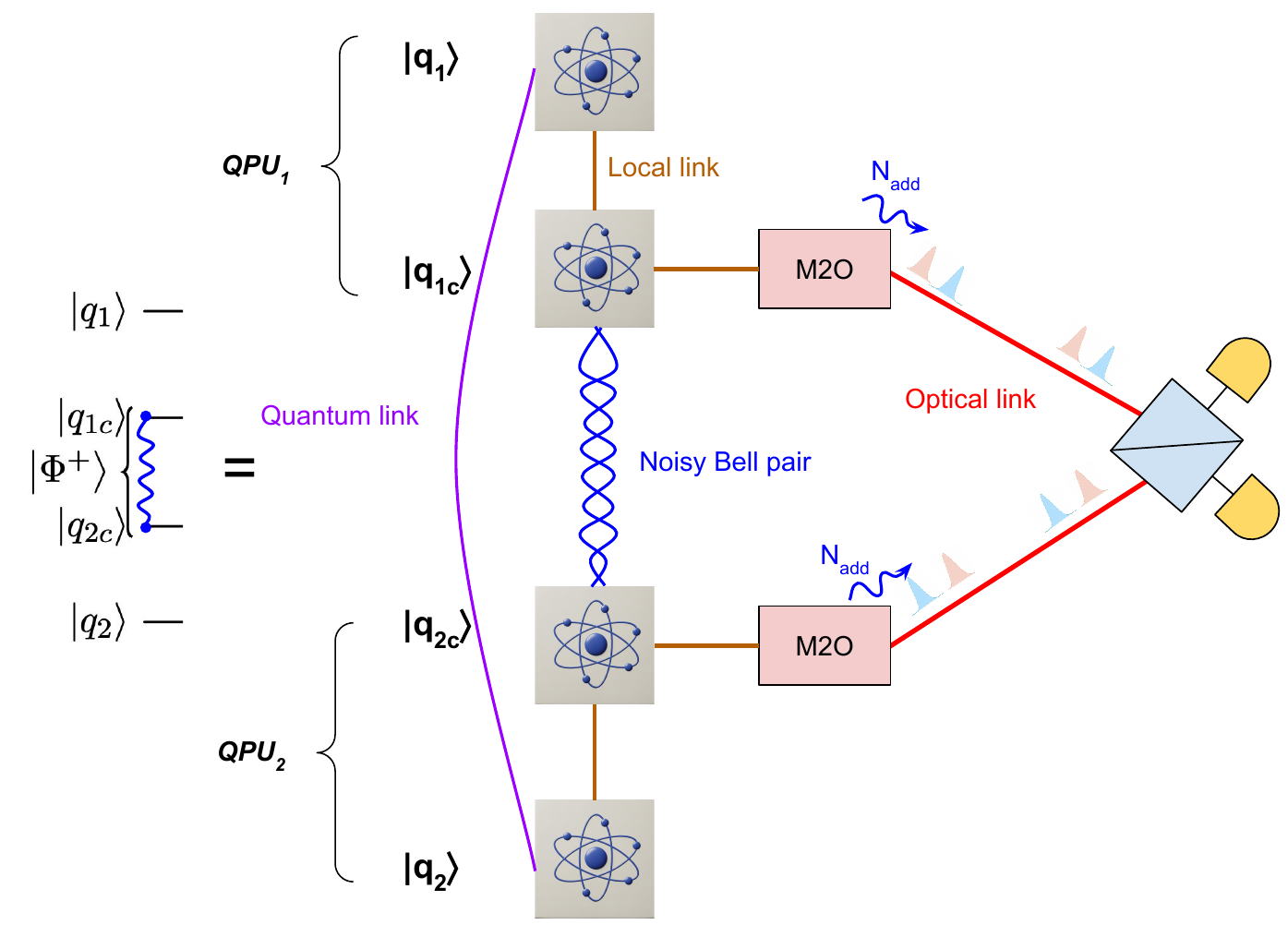}
    \caption{Quantum link between qubits $\ket{q_1}$ and $\ket{q_2}$ in two different modules is established using communication qubits $\ket{q_{1c}}$ and $\ket{q_{2c}}$ and microwave-to-optical (M2O) transducers over an optical link. The qubits $\ket{q_{1c}}$ and $\ket{q_{2c}}$ share a noisy Bell pair, with the primary source of infidelity being the noise added by transducer $N_{add}$. A remote 2-qubit gate is executed between data qubits $\ket{q_1}$ and $\ket{q_2}$ through a combination of local gates and a Bell pair between the communication qubits. }
    \label{fig:3}
\end{figure}

To implement remote gates, we employ the TeleGate primitive \cite{Van_Meter2016-gs,Cuomo2020-qb} (Fig.~\ref{fig:2}), which uses a shared Bell pair between two communication qubits across the modules along with local operations to implement teleported gates. Fig.~\ref{fig:3} shows the physical implementation of the optical quantum link between the two modules. Each data qubit $\ket{q_1}$ and $\ket{q_2}$ has a corresponding communication qubit $\ket{q_{1c}}$ and $\ket{q_{2c}}$  connected via intramodule links. The communication qubits are connected to microwave-to-optical transducers, which upconvert microwave photons to optical photons, which are then routed over a room-temperature optical fiber link to a Bell state measurement (BSM) setup. Entanglement between communication qubits is optically heralded, and subsequently local operations are performed to implement remote gates between data qubits $\ket{q_1}$ and $\ket{q_2}$ across modules. 

The remote 2-qubit gate is expected to be noisier than local gates due to the introduction of noise photons by M2O transducers, the optical channel, and the need for additional 1-qubit, 2-qubit gates, and measurements to execute the teleported gate (Fig.~\ref{fig:2}). Inside a datacenter scheme, where optical channel length is $\sim$10-100 m, the transducer performance is expected to be the primary bottleneck for achieving high-rate, high-fidelity entanglement between superconducting qubits. The transducer can be described by a few key metrics, which include the end-to-end microwave-to-optical upconversion efficiency $\eta$, bandwidth $B$, operation time $T$ and the input-referred noise added $N_{add}$. Among these parameters, noise-added is of primary interest for evaluating entanglement fidelity \cite{zeuthen2020,PhysRevX.12.021062}, with $N_{add} <$ 1 required for photonic entanglement of superconducting qubits. We can describe the dark count probability due to added noise  $P_d$, by the formula $P_d = \left(1 - e^{-\frac{r_NT}{2}} \right)^2$, where $r_N =  \eta B N_{add} $ is the added noise rate at the optical output of the transducer  \cite{dirnegger2025montecarlomodeldistilled,zeuthen2020}.

We study the two-click heralded entanglement generation protocol \cite{PhysRevA.71.060310} because of its higher robustness to noise over the one-click protocol, with some trade-off in rates due to a quadratic dependence on transducer efficiency \cite{dirnegger2025montecarlomodeldistilled}. The entanglement fidelity for the two-click protocol primarily depends on the noise added $N_{add}$ and does not show a significant dependence on the transducer efficiency \cite{zeuthen2020,dirnegger2025montecarlomodeldistilled}. The density matrix of a noisy Bell pair that we inject in the TeleGate protocol is given by the formula \cite{dirnegger2025montecarlomodeldistilled,zeuthen2020}

\begin{align*}
    \label{BellDensity}
    \sigma &= \frac{1}{N}\{(1-P_e^2)2P_d(1-P_d)([1-(1-\eta)^2](1-P_d) \\  &+ (1-\eta)^2 2P_d (1-P_d))\ket{00}\bra{00}\\
    & + 2P_e(1-P_e) \eta^2 (1-P_d)^2 \ket{\Psi^+} \bra{\Psi^+} \\
    & + ((\eta(1-P_d)+(1-\eta)(1-P_d)2P_d)^2  \\ & - \eta^2 (1-P_d)^2)(\ket{01}\bra{01} + \ket{10}\bra{10}) \\
    & + P_e^2 (( 1- (1-\eta)^2 ) + (1-\eta)^2 2 P_d) (1-P_d)^2 2 P_d \ket{11} \bra{11}\}
\end{align*}

\noindent
where $N$ is a normalizing factor and $P_e$ is excitation probability, kept as 0.5, which is optimum for two-click entanglement generation protocols \cite{dirnegger2025montecarlomodeldistilled}. We fix the transducer bandwidth B to 10 MHz, operation time T to 1 $\mu s$ based on comparable values reported for integrated electro-opto-mechanical quantum transducers \cite{weaver2025scalablequantumcomputingoptical,Weaver2024-uh,Jiang2023-zk,Meesala2024-mk,Warner2025-sm}. The efficiency was fixed to 0.5, which is the highest conversion efficiency reported for a 3D electro-opto-mechanical device \cite{PhysRevX.12.021062}. We note that the Bell pair fidelity is expected to exhibit a weak dependence on the transducer efficiency $\eta$ \cite{zeuthen2020} and therefore in our analysis we vary the noise added values $N_{add}$ to evaluate the remote gate fidelity.

% \begin{figure}[!htpb]
%     \centering
%         \begin{quantikz}[color=blue]
%         \lstick{\ket{q_1}}  & \ctrl{1}  &  \\
%         \lstick{\ket{q_2}}  & \targ{} &
%     \end{quantikz}=\begin{quantikz}
%         \lstick{\ket{q_1}} & \ctrl{1} & & & & \gate{Z} & \\
%         \lstick[2]{\ket{\Phi^+}}  \lstick{\ket{q_{1c}}} & \targ{} & \meter{} \wire[d][1]{c} \\
%         \lstick{\ket{q_{2c}}} & & \gate{X} & \ctrl{1} \gategroup[2, steps = 1, style = {dashed, inner sep=6pt}, label style = {label position = below, anchor = north, yshift = -0.2cm}]{Teleported Gate}& \gate{H} & \meter{} \wire[u][2]{c} \\
%         \lstick{\ket{q_2}} &&& \targ{} &&&
%     \end{quantikz}
%     \caption{Example of a remote CNOT gate.}
%     \label{fig:tcanther}
% \end{figure}

% \begin{figure}[!htpb]
%     \centering
%     \begin{quantikz}[color=blue]
%         \lstick{\ket{q_1}}  & \ctrl{1}  &  \\
%         \lstick{\ket{q_2}}  & \targ{} &
%     \end{quantikz}
%     \caption{Example of a remote CNOT gate.}
%     \label{fig:tcanther}
% \end{figure}

% \subsubsection{Physical model}

% \subsubsection{Injecting noisy Bell pair}
To simulate noisy remote gates in Qiskit AER \cite{qiskit2024}, we apply a noisy Bell pair state preparation circuit to the communication qubits. We start with a CNOT gate between the communication qubits to generate Bell state $\ket{\Phi^+} = (\ket{00} + \ket{11})/\sqrt{2}$, followed by $X \otimes I$ operation to map it to state $\ket{\Psi^+} = (\ket{01} + \ket{10})/\sqrt{2}$. To map this ideal Bell pair density matrix $\rho = \ket{\Psi^+} \bra{\Psi^+}$ to the noisy Bell pair matrix $\sigma$ described above, we use a trivial replacement channel \[\mathcal{E}_{\sigma}(\rho) = \mathrm{Tr}(\rho)\sigma=  \sum_{i,j}K_{i,j}\rho K_{i,j}^{\dagger}\] (see Appendix C on Kraus operators $K_{i,j}$). This replacement channel is added as a custom noise channel in Qiskit to the CNOT gate between the designated communication qubits during the noisy Bell state preparation step. 

% Local single qubit, two qubit and measurement errors are all set to 0.05. 

\begin{figure}
    \centering
    \includegraphics[width=0.9 \linewidth]{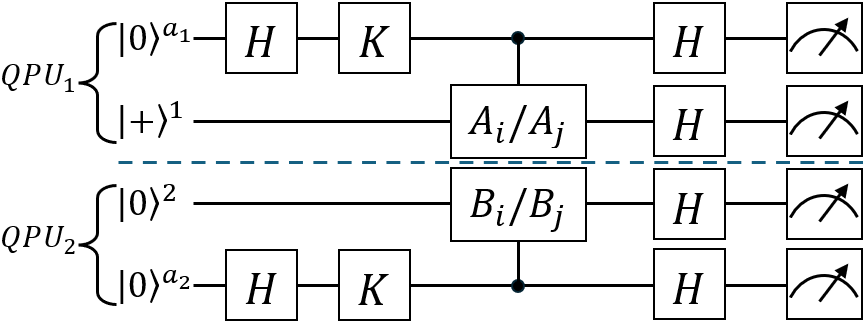}
    \caption{Illustration of the quasiprobability-based circuit cutting procedure applied to the CNOT gate connecting two modules. The measurement of the observable $X \otimes X$ is performed on data qubits 1 and 2. The ancilla qubits, $a_1$ and $a_2$, mediate the measurement of state overlaps for cases where $i \neq j$.}
    \label{Fig:CNOT_QPD}
\end{figure}

\subsection{Circuit cutting}
Circuit cutting \cite{Peng_2020, Perlin_2021, Liu_2022, Basu_2024, Chen_2024, ayral2020quantum} is a method that lets us run larger quantum circuits than current devices can directly handle. The idea is to split a circuit into smaller, hardware-compatible pieces by cutting either qubit wires or multi-qubit gates. Each piece, or sub-circuit, can be executed separately, and the results are combined through classical post-processing to reconstruct the outcome of the original circuit. While this comes with additional sampling overhead, circuit cutting makes it possible to study problems that would otherwise exceed the limits of today’s quantum processors and even helps avoid costly swap operations between distant qubits.

There are two main types of cuts. Wire cuts (time-like cuts) separate a qubit line, replacing it with a measurement followed later by a corresponding state preparation. Gate cuts (space-like cuts) decompose a multi-qubit gate into a weighted set of simpler, local operations—a process described by a quasiprobability distribution (QPD). The Qiskit circuit-cutting-addon library \cite{circuit_knitting_toolbox} provides practical tools for both approaches, and in our work, we focus on gate cutting, applying it to GHZ-type circuits.

Let us consider a unitary that can be decomposed into a sum of local operations on two quantum processing units (QPUs), 1 and 2, such that
\begin{align} \label{Eq:Unitary_Decomposition}
U = \sum_i \alpha_i A_i^{(1)} \otimes B_i^{(2)},
\end{align}
where the coefficients \(\alpha_i\) may take negative values. To evaluate the expectation value of an observable $\mathcal{O} = P^{(1)} \otimes Q^{(2)}$ with respect to the product state $\rho = \sigma^{(1)} \otimes \tau^{(2)}$, we must perform a sufficient number of local experiments to accurately approximate
\begin{align} \label{Eq:Expectation_Value}
    \text{tr}(\mathcal{O} \cdot U \rho U^\dagger) = \sum_{i,j} \alpha_i \alpha_j \text{tr}(P A_i \sigma A_j^\dagger) \text{tr}(Q B_i \tau B_j^\dagger).
\end{align}
The Hadamard test \cite{luongo2024hadamard} provides a method to locally compute the overlaps ($i \neq j$) appearing in the trace terms of Eq.~\ref{Eq:Expectation_Value}, allowing the expectation value to be reconstructed from measurements on the individual QPUs. If additional ancilla qubits are not available on the quantum processor, the local overlaps can instead be computed by exploiting the cyclic property of the trace. In this case, it is sufficient to evaluate only the local expectation values $\text{tr}(A_j^\dagger P A_i \sigma)$ on QPU 1 and $\text{tr}(B_j^\dagger Q B_i \tau)$ on QPU 2; however, there will be additional overhead that slows down computation time.

To calculate the expectation value using a quasiprobability distribution (QPD), we define probabilities $p_k = |\alpha_k| / \sum_i |\alpha_i|$ and assign them the weights $\sum_i |\alpha_i| \, \text{sgn}(\alpha_k)$. This gives
\begin{align} \nonumber
    &\sum_{k,l} |\alpha_k| |\alpha_l| \\ \times &\sum_{i,j} p_i p_j \, \text{sgn}(\alpha_i) \text{sgn}(\alpha_j) \text{tr}(P A_i \sigma A_j^\dagger) \text{tr}(Q B_i \tau B_j^\dagger),
\end{align}
which can be evaluated locally on each QPU by running a sufficient number of shots for each subexperiment. This approach ensures that the expectation value of the original global operator is faithfully reconstructed from local measurements. Qiskit's circuit cutting library will automatically implement the QPD after identifying the cut points for the gates acting across the QPUs.

We use Figure \ref{Fig:CNOT_QPD} as an example of how the QPD can be performed in practice on the CNOT gate across qubits 1 and 2. Let's assume we prepare the Bell state $\ket{\Phi^+} = (1/\sqrt{2})(\ket{00} + \ket{11})$ and want to obtain the expectation value $\langle X \otimes X \rangle_{\Phi^+}$ for the observable $X \otimes X$. For simplicity, we write the CNOT gate as
\begin{align}
    CNOT = \dfrac{1}{2}(\mathbb{I} \otimes \mathbb{I} + \mathbb{I} \otimes X + Z \otimes \mathbb{I} + [-Z] \otimes X),
\end{align}
which has the desired form in Eq. \ref{Eq:Unitary_Decomposition}. For instance, we can define $A_1 = A_2 = \mathbb{I}$, and $A_3 = -A_4 = Z$, $B_1 = B_3 = \mathbb{I}$, and $B_2 = B_3 = X$. We also define the controlled gates in Fig. \ref{Fig:CNOT_QPD} as
\begin{align}
   C(A_i, A_j) &= \ket{0} \bra{0} \otimes A_i + \ket{1} \bra{1} \otimes A_j \\
    \text{and } C(B_i, B_j) &= \ket{0} \bra{0} \otimes B_i + \ket{1} \bra{1} \otimes B_j.
\end{align}
In \cite{Harrow_2025} it was shown that the double Hadamard test can be used to perform the QPD to obtain the overlap values when $i \neq j$. So we introduced ancillary qubits $a_1$ and $a_2$ to perform this task.

To perform the QPD properly, we must sample according to the probability mass function
\begin{align}
    p(i, j, k) = \begin{cases}
        0 \quad &\text{if } i = j \text{ and } k = 1 \\
        \alpha_i \alpha_j q(U)^{-1} \quad &\text{otherwise}
    \end{cases}
\end{align}
where $q(U) = 2||\vec{\alpha}||_1^2 - ||\vec{\alpha}||_2^2$ is directly related to the sampling cost of this distribution and $||\star||_k$ is the $k$-norm of the vector of coefficients $\vec{\alpha} = \{ \alpha_1, \alpha_2, ..., \alpha_n \}$. Specifically, to estimate the expectation value of the original circuit, using this particular strategy with the double Hadamard test, the number of samples required is at most on the order of $q(U)^2 ||\mathcal{O}||^2$ for the observable $\mathcal{O}$ \cite{Harrow_2025}. Therefore, the goal would be to minimize the value $q(U)$ over all decompositions of $U$. 

In our work, we apply gate cutting to a variety of GHZ-type circuits. We cannot use the Qiskit circuit-cutting-addon library directly to calculate the counts because it only supports obtaining the expectation values of an observable. To get the counts, we must run $2^n$ experiments for $n$-qubit systems, which drastically increases the resources necessary for calculating the Hellinger fidelity. To illustrate how this can be done, we use a two-qubit example for simplicity. The reconstructed expectation values are given by
\begin{align}
    E_{00} &= \text{tr} [(\mathbb{I} \otimes \mathbb{I}) \cdot \rho^{12}] \qquad \; E_{01} = \text{tr} [(\mathbb{I} \otimes Z) \cdot \rho^{12}] \\
    E_{10} &= \text{tr} [(Z \otimes \mathbb{I}) \cdot \rho^{12}] \qquad E_{11} = \text{tr} [(Z \otimes Z) \cdot \rho^{12}],
\end{align}
which are enough to derive the probabilities
\begin{align}
    p_{00} &= \dfrac{1}{2}(E_{00} + E_{01} + E_{10} + E_{11}) = \text{tr} [\ket{00} \bra{00} \cdot \rho^{12}] \\
    p_{01} &= \dfrac{1}{2}(E_{00} - E_{01} + E_{10} - E_{11}) = \text{tr} [\ket{01} \bra{01} \cdot \rho^{12}] \\
    p_{10} &= \dfrac{1}{2}(E_{00} + E_{01} - E_{10} - E_{11}) = \text{tr} [\ket{10} \bra{10} \cdot \rho^{12}] \\
    p_{11} &= \dfrac{1}{2}(E_{00} - E_{01} - E_{10} + E_{11}) = \text{tr} [\ket{11} \bra{11} \cdot \rho^{12}].
\end{align}
This can be expressed nicely by the equation $\vec{p} = (H \otimes H) \cdot \vec{E}$, where $H$ is the Hadamard gate, $\vec{p}$ is the column vector $\{p_{00}, p_{01}, p_{10}, p_{11} \}^T$, and $\vec{E}$ is the column vector $\{E_{00}, E_{01}, E_{10}, E_{11}\}^T$. For $n$ qubits, the general formula is given by the Walsh-Hadamard transform \cite{Walsh1923, ShuklaVedula2022} $\vec{p} = H^{\otimes n} \cdot \vec{E}$, where the subscripts in vectors $\vec{p}$ and $\vec{E}$ are ordered from smallest to largest in binary; that is, for a three-qubit system we have that 
\begin{align}
    \vec{p} = \{ p_{000}, p_{001}, p_{010}, p_{011}, p_{100}, p_{101}, p_{110}, p_{111} \}^T.
\end{align}
If we can derive $\vec{p}$ from the reconstructed vector $\vec{E}$, and if we know the total number of counts $N$, we can estimate the number of counts for each set of outputs as $N \cdot \vec{p}$. 

This work focuses exclusively on local operations (LO), without incorporating classical communication, in our comparison of results. While local operations and classical communication (LOCC) can be employed to further reduce circuit-cutting overhead—yielding $\gamma = 4$ for LOCC compared to $\gamma =9$ for LO—our analysis isolates the effects of purely local strategies. The exclusion of classical communication allows for a more direct assessment of the intrinsic quantum performance and scalability of the approach, independent of classical coordination overhead. In practice, real-time classical communication between quantum modules can be integrated with circuit-cutting techniques to reduce overall runtime overhead \cite{Carrera_Vazquez2024-gy,Piveteau2024-bi}.

% Results
\section{Results}

\begin{figure}[!htpb]
    \centering
    \includegraphics[width=0.99\linewidth]{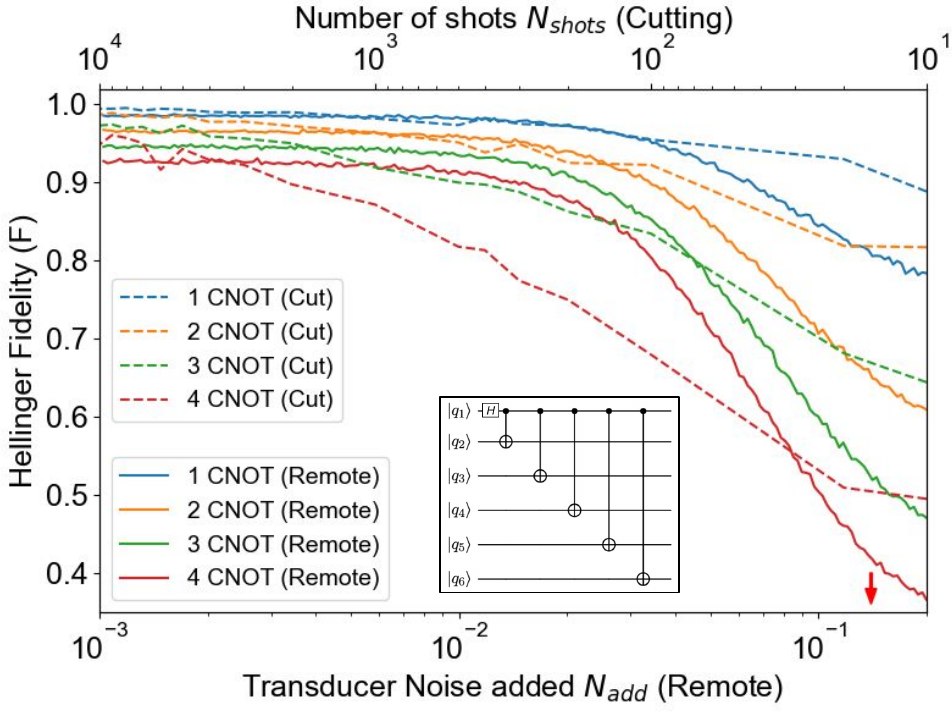}
    \caption{GHZ Hellinger Fidelity with remote CNOT gates (solid lines) vs transducer noise added (${N_{add}}$) on the bottom x-axis and CNOT gate-cuts (dashes) vs number of shots (${N_{shot}}$) on the top x-axis. Remote gate fidelity shows a strong dependence on the transducer noise added, and the gate-cut fidelity depends on the number of shots. In the regime of low ${N_{add}}$ (remote) and high ${N_{shots}}$ (cut), the local gate errors dominate the GHZ fidelity. Red arrow indicates state-of-the-art transducer ${N_{add}}$ figure \cite{Meesala2024-mk,Warner2025-sm}.}
    \label{fig:6}
\end{figure}

Fig.~\ref{fig:6} shows the Hellinger fidelity of GHZ circuits of size N (2-5) created using remote CNOT gates in solid lines and CNOT gate-cuts in dashed lines. The remote gate GHZ fidelity is plotted as a function of the transducer noise added (${N_{add}}$), as labeled in the bottom x-axis. The fidelity shows an initial weak dependence on the noise, followed by a strong suppression for ${N_{add}} \gtrsim$0.04, with a higher fidelity degradation for GHZ circuits of larger sizes. The fidelity in the low ${N_{add}}$ regime ($\lesssim0.01$) is limited by the local two-qubit gate fidelity of 0.98 instead of the transducer noise. 

The gate-cut fidelity is plotted as a function of the repetitions of each sub-circuit ($N_{shots}$), as labeled in the top x-axis of Fig.~\ref{fig:6}. The fidelity shows degradation with decreasing $N_{shots}$, as well as with increasing number of gates N for a fixed $N_{shots}$. An insufficient sampling of the quasiprobability decomposition of the gate cut can explain the trend. For the case of N CNOT gate-cuts, the $N_{shots}$ required to reconstruct the original circuits with a reconstruction error $\epsilon$ is expected to scale as $O(9^{N}/\epsilon^2)$, which can be attributed to Hoeffding's inequality \cite{PhysRevLett.125.150504,yang2024understandingscalabilitycircuitcutting} (see Appendix B). For example, in Fig.~\ref{fig:6}, $N_{shots}$ required to maintain a fixed fidelity F=0.9 in Fig.~\ref{fig:6} is $\sim$10, 70, 520, and 2100 for 1, 2, 3, and 4 CNOT gate-cuts, respectively. For the largest number of $N_{shots} =$10000, the highest fidelity is F= 0.994, 0.986, 0.973, and 0.948 for N= 1, 2, 3, and 4 gate-cuts, respectively, limited by local two-qubit gate fidelity of 0.98 and the shot budget.

We observe two distinct regimes in Fig.~\ref{fig:6}, the low $N_{add}$/high $N_{shots}$ regime, where local 2-qubit error dominates, and another intermediate regime where the error is dominated by transducer pair noise for remote gates and sampling error for gate-cuts. In the first regime (left edge of Fig.~\ref{fig:6}), the gate cutting fidelity outperforms remote CNOT fidelity for all numbers of CNOT gates, N. This is expected since the teleported CNOT gate incurs a higher local error from additional 1-qubit and 2-qubit gates required by the telegate protocol (Fig.~\ref{fig:2}). In the second regime, for a limited $N_{shots}$, remote gates can provide higher fidelity than gate cuts. The crossover between the two regimes in Fig.~\ref{fig:6} happens for $N_{shots} \sim$600, 1000, 1800, 4000 for 1, 2, 3 and 4 CNOT gates, respectively. Thus, for generating multipartite entangled states, circuit cutting requires progressively larger numbers of shots to outperform remote gates. As the local two-qubit gate error decreases, we anticipate that the crossover $N_{shots}$ will increase because of the reduction in the additional local gate penalty for remote gates. 

\begin{figure}[!htpb]
    \centering
    \includegraphics[width=0.99\linewidth]{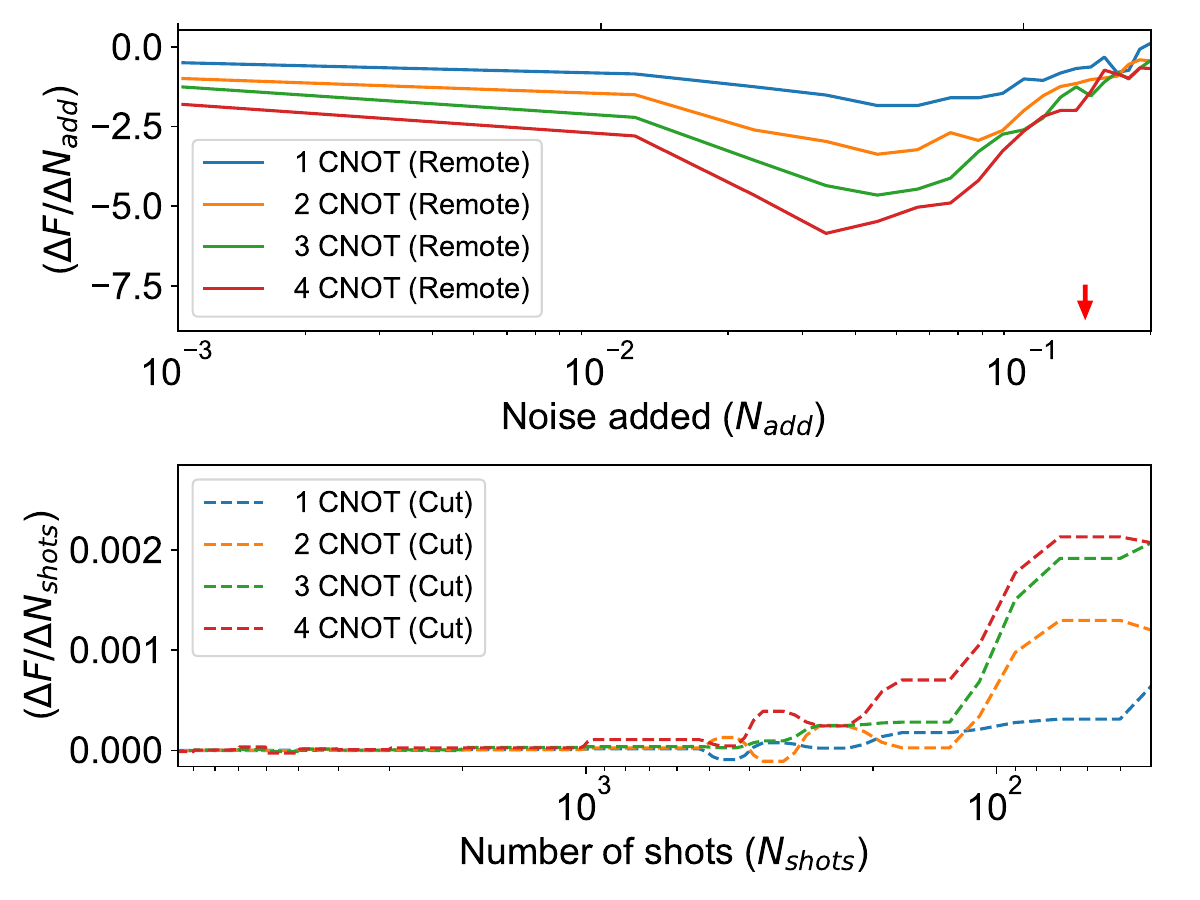}
    \caption{Derivative of fidelity of remote CNOT gates vs transducer noise added, ${ {\Delta F} /{\Delta N_{add}}}$ (top panel) and fidelity of CNOT gate-cuts vs number of shots, ${{\Delta F}/{\Delta N_{shot}}}$ (bottom panel). ${{\Delta F}/{\Delta N_{add}}}$ has largest magnitude in the regime of ${N_{add}}$ 0.01-0.1, with red arrow indicating the state-of-the-art ${N_{add}} \sim $ 0.1 \cite{Meesala2024-mk,Warner2025-sm}.  The derivative of circuit cutting fidelity  ${{\Delta F}/{\Delta N_{shot}}}$ has the largest magnitude for ${N_{shots}}$ \textless $\sim$ 1000.}
    \label{fig:7}
\end{figure}

Fig.~\ref{fig:7} shows the derivative of the GHZ fidelity vs ${N_{add}}$ for remote gates (${ {\Delta F} /{\Delta N_{add}}}$) in the top panel. The fidelity derivative approaches zero in two regimes, a high noise-added regime for ${N_{add}} \gtrsim$0.2 and a very low noise-added regime for ${N_{add}} \lesssim$0.001. Between these two regimes, the fidelity derivative has a peak between ${N_{add}} \sim$0.01-0.1, suggesting that a reduction in transducer noise in this regime will provide the largest improvement in GHZ state fidelity.

The bottom panel in Fig.~\ref{fig:7} shows the derivative of the GHZ fidelity vs number of shots ${{\Delta F}/{\Delta N_{shot}}}$. Fidelity derivative shows suppression between ${N_{shots}} =$ 100-1000 for GHZ circuit sizes, indicating that a shot budget of 1000 is necessary for optimum fidelity.
The magnitude of the derivative of fidelity is larger with larger GHZ circuit sizes (number of CNOT gates) in both cases, indicating that larger entangled states are more susceptible to accumulation of errors in both cases, i.e., gate-cuts and remote gates. 

\begin{figure}[!htpb]
    \centering
    \includegraphics[width=0.99\linewidth]{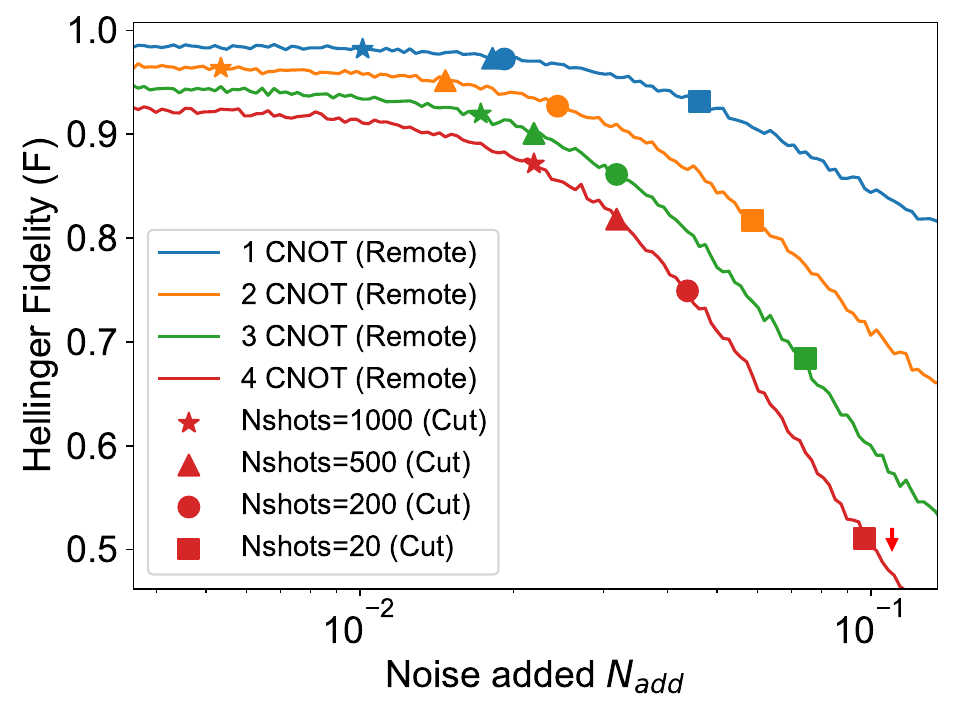}
    \caption{Hellinger Fidelity of remote CNOT gates (solid lines), with scatter points highlighting the threshold ${N_{add}}$ where the circuit cutting fidelity at select shot budget ${N_{shot}}$ matches the remote gate fidelity. Stars (${N_{shot}}$=1000), Triangles (${N_{shot}}$=500), Circles (${N_{shot}}$=200) and Square (${N_{shot}}$=20). Line plots and scatter plots with the same color correspond to the same number of CNOT gates. Red arrow indicates state-of-the-art transducer ${N_{add}}$ \cite{Meesala2024-mk,Warner2025-sm}. }
    \label{fig:8}
\end{figure}

\begin{figure}[!htpb]
    \centering
    \includegraphics[width=0.99\linewidth]{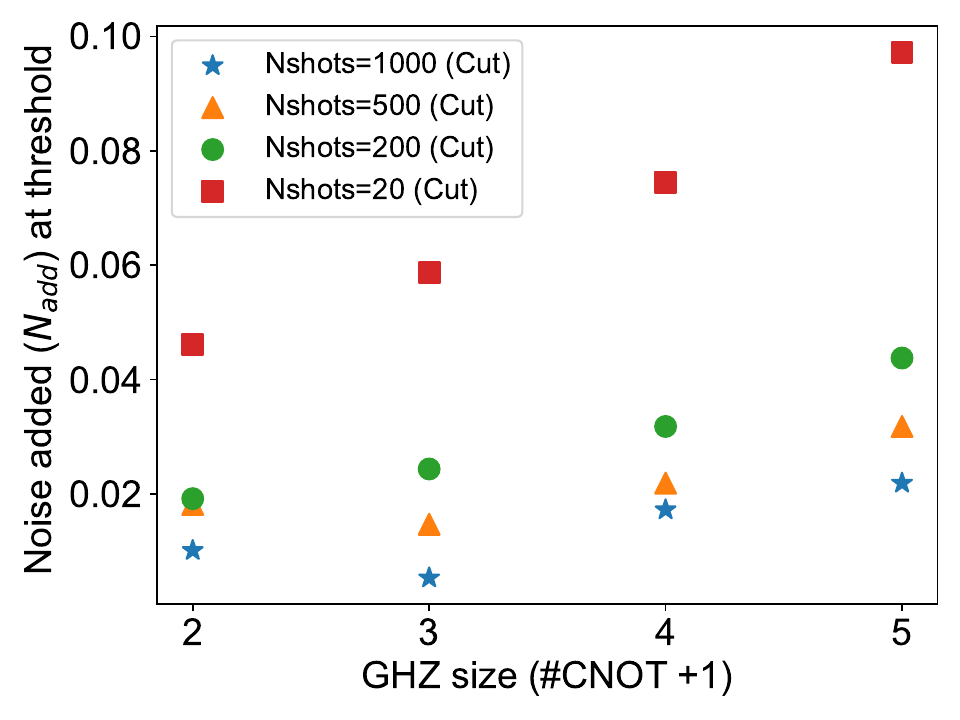}
    \caption{Transducer noise added (${N_{add}}$) at the threshold vs GHZ circuit size at different values of circuit-cutting shot budget ${N_{shot}}$. At a smaller shot budget, the circuit cutting fidelity is degraded, resulting in a higher tolerance for the quantum link noise at the threshold point. The ${N_{add}}$ at threshold shows an increase with increasing GHZ circuit siz,e suggesting that remote gates are advantageous for generating multipartite entangled states.}
    \label{fig:9}
\end{figure}

Fig.~\ref{fig:8} plots the remote gate fidelity vs ${N_{add}}$ (lines), with points where the remote gate fidelity matches the gate-cut fidelity for a fixed shot budget plotted as scatter points ($F_{Remote}(N_{add})=F_{cut}(N_{shot})$). For ${N_{shots}}$ of 1000 (star), 500 (triangle), 200 (circle), and 20 (square), the threshold value of ${N_{add}}$ increases from 0.01 to 0.1. Fig.~\ref{fig:9} plots this ${N_{add}}$ threshold as a function of GHZ circuit size, for the same values of shot budget as Fig.~\ref{fig:8}. We observe a slight increase in ${N_{add}}$ threshold with increasing GHZ circuit sizes. This suggests that while both approaches, i.e., remote gates and gate-cuts, suffer from a reduction in fidelity for larger GHZ state sizes, the penalty for gate-cuts is larger than for remote gates, which leads to an increase in ${N_{add}}$ threshold. Therefore, with a limited shot budget and larger entangled states (more distributed CNOT gates), the break-even point for transducer hardware noise is relaxed, such that the two approaches can have comparable fidelity. We attribute these trends in ${N_{add}}$ threshold to the exponential sampling overhead for circuit-cutting, which leads to a reduction in fidelity at a fixed shot budget ${N_{shot}}$ with a larger number of gate-cuts. The threshold noise values in Fig.~\ref{fig:9} are between 0.01 and 0.1, which is within 10-fold of the state-of-the-art transducer noise figure of $\sim$0.1 \cite{Meesala2024-mk,Warner2025-sm}.

% Discussion
\section{Discussion}

Our results suggest that, within a 10-fold reduction in transducer noise over the current experimental state-of-the-art, remote gate fidelity can approach that of circuit-cutting-based approaches for generating intermodule GHZ states. This motivates us to pursue a reduction in total quantum run time (shot budget) for distributed computation by utilizing quantum links of sufficient fidelity. However, since near-term quantum links will be probabilistic, the key source of latency in remote gate speeds is the waiting time between successful entanglement attempts. We describe a network-aware algorithm (Fig.~\ref{code}) which dynamically chooses gate cuts or remote gates based on the availability of Bell pairs. All non-local gates start with a cut, with the total shot budget equally divided across all gates. For each link i, the fidelity for gate cut at the given $N_{shots}$ is compared with the fixed fidelity of the quantum link. If the quantum link offers higher fidelity and if a Bell pair is available, the algorithm chooses remote gates. Alternatively, if gate-cut offers higher fidelity or if quantum links are not available, the algorithm decides gate-cuts by increasing the shot budget $S_i$ for the particular link i to $S_i'$, when necessary.

\begin{algorithm}
\begin{algorithmic}

\State $S_0 \gets S_{total}$
\ForAll{$i \in \{1, \dots, N\}$}
    \State $S_i \gets S_0/(N-(i-1))$
    \If{$ F_{cut}(S_i) \geq F_{remote}^i$} 
        $Gate \,\, cut$
        \State $S_0 \gets S_0 - S_i$ 
    \ElsIf{$Bell \,\, pair  \,\, available$}
            $Remote \,\, gate$
    \Else
        \State $S_i \gets S_i' $, {\bf and} $Gate \,\, cut$
        \State $S_0 \gets S_0 - S_i'$

    \EndIf 
\EndFor 
\label{code}
\end{algorithmic}
\caption{A network-aware greedy algorithm for choosing between gate cuts and remote gates}
\end{algorithm}

% Conclusion
\section{Outlook}

Our current work demonstrates utility for GHZ state generation with remote gates, which will be expanded to evaluate other algorithms and subroutines for their suitability in circuit-cutting versus remote gates. This includes algorithms in variational quantum circuits \cite{sünkel2025evaluatingvariationalquantumcircuit}, quantum chemistry \cite{Jones_2024}, and quantum machine learning \cite{Neumann2022-sc,Pira2023-jy}. 

% \begin{figure}[!htpb]
%     \centering
%     \includegraphics[width=0.99\linewidth]{Nikolay report/algo_simplifield.pdf}
%     \caption{Greedy algorithm for optimizing total shot budget $S_0$ given N gates between QPU.} 
%     \label{fig:5}
% \end{figure}

We will build on the greedy algorithm outlined in section IV and develop a hybrid distributed compiler based on a unified cost function that incorporates the space-time quantum overhead associated with the two approaches, as well as classical post-processing overhead for circuit-cutting. For example, for the time overhead,  CNOT gate-cuts have a quantum runtime scaling as $\sim 9^N$ for N CNOT gates, and the corresponding time latency for generating N Bell pairs for N remote gates scales as $\sim 1/\eta^{2N}$. For an approximate latency/runtime break-even point between the two approaches, we have $9^N = 1/\eta^{2N}$, which requires transducer efficiency $\eta > 0.33$. A distributed circuit compiler can minimize the Bell pair budget \cite{Ferrari_2023,mengoni2025efficientgatereorderingdistributed} , which will reduce the break-even efficiency for near-term transducers. While transducer efficiencies as high as 0.47 have already been reported for a 3D transducer, the device exhibited an $N_{add}$ of 3.2 \cite{PhysRevX.12.021062}, significantly higher than the threshold of $N_{add}$ \textless\ 0.1 discussed in our work. Scalability and integration consideration ideally requires chip-scale transducers, which exhibit a lower efficiency of $\sim$1 \% at a similar $N_{add}$ of 3.2 \cite{Xie2025-ch}. Therefore, a key challenge will be maximizing the transducer efficiency while keeping the noise added due to the optical pump below the threshold \cite{weaver2025scalablequantumcomputingoptical}, and our simulation effort will inform on the co-optimization of the transducer hardware metrics from an application point of view. We will further explore this remote gate fidelity-rate trade-off by modeling different transducer pumping conditions, incorporating entanglement distillation protocols \cite{PhysRevX.4.041041,weaver2025scalablequantumcomputingoptical,dirnegger2025montecarlomodeldistilled}, as well as error-mitigation techniques for quantum links \cite{Ang2024-fh}.  The remote gate noise model will be expanded to include other sources of errors, including dephasing and control errors on communication qubits during Bell pair generation, dephasing of data qubits, noise photons, and depolarization in the optical fiber. Our noise model is general for remote gates based on the two-click protocols for other qubit modalities that are interconnected with noisy frequency converters and will be expanded in future works to include other qubit-specific noise sources. The entanglement generation protocol will also be expanded to include the single-click protocol, optical-to-microwave (O2M) downconversion protocol, and spontaneous parametric downconversion (SPDC) source-based protocols \cite{Ang2024-fh}.

% Conclusion
\section{Conclusion}

We perform a simulation of intermodule GHZ states generated for superconducting qubits using remote gates over noisy optical links. We inject the effect of noise added by microwave-to-optical transducer noise $\mathrm{N_{add}}$ into entanglement-based remote gates and perform comparative simulations of remote gates (quantum) with gate-cuts (classical). We observe a degradation in remote gate fidelity with increasing $\mathrm{N_{add}}$, as well as with increasing GHZ circuit size for fixed $\mathrm{N_{add}}$. The gate-cut fidelity on the other hand, depends on the number of repetitions of sub-circuits ($\mathrm{N_{shot}}$) and shows a similar degradation with increasing GHZ circuit size (N) due to an exponential increase in sampling overhead associated with circuit cutting techniques. As a result, for a fixed shot budget, we find that remote gates can exhibit comparable fidelity to gate-cuts if the added transducer noise is below a threshold. The noise threshold increases with larger GHZ circuit sizes, making remote gates preferable for generating multipartite entangled states over gate-cuts. The corresponding $\mathrm{N_{add}}$ threshold of (0.01, 0.1) is within an order of magnitude of the best noise figure ($\sim$0.1) for state-of-the-art transducers \cite{Meesala2024-mk,Warner2025-sm}. The derivative of the GHZ fidelity with transducer noise also shows the most substantial dependence on $\mathrm{N_{add}}$ in the regime (0.01, 0.1); therefore, a 10-fold reduction in current noise figures of transducers will result in a significant improvement in fidelity of multipartite entangled states. With these improvements, near-term optical links can provide comparable fidelity as circuit-cutting techniques, making the case for hybrid classical-quantum computation, which leverages sparse quantum links with circuit-cutting techniques.

% Conclusion
\section{Appendix}

\subsection{Hellinger Fidelity}

To measure the accuracy of a noisy simulation, we use Hellinger Fidelity. It is a quantity measuring the closeness between two probability distributions. For discrete distributions $P = (p_1, \ldots, p_k)$ and $Q = (q_1, \ldots, q_k)$ it is given by \[H = \left(\sum_i \sqrt{p_i q_i}\right)^2.\] Hellinger fidelity coincides with quantum fidelity for diagonal density matrices; however, it can differ significantly in the presence of big enough off-diagonal elements.

\subsection{Variance}
From gate cutting, we decompose the gate as
\begin{align}
    \mathcal{G}(\rho)=\sum_ic_i\mathcal{K}_i(\rho).
\end{align}
Then the sample expectation $\mathbb{E}_{\hat O}$ value of an observable $O$ is
\begin{align}
    \hat\mu=\mathbb{E}_{\hat O}[\mathcal{G}(\rho)]=\sum_ic_i\mathbb{E}_{\hat O}[\mathcal{K}_i(\rho)].
\end{align}
The sample variance is
\begin{align}
    \text{var}(\hat\mu)=\text{var}(\sum_ic_i\mathbb{E}_{\hat O}[\mathcal{K}_i(\rho)])=\sum_ic_i^2\text{var}(\mathbb{E}_{\hat O}[\mathcal{K}_i(\rho)]),
\end{align}
where the variance is distributed because the sample expectation values for the fragments are independent. Therefore, for multiple cuts, the variance scales exponentially with the number of cuts.

\subsection{Replacement channel for noisy remote gates}

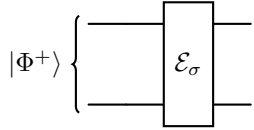
\begin{figure}[!htpb]
    \centering
    \begin{quantikz}
        \lstick[2]{\ket{\Phi^+}} & & \gate[2]{\mathcal{E}_{\sigma}} &  \\
        &&&
    \end{quantikz}
    \caption{Injecting a noisy Bell pair with density matrix $\sigma$}
    \label{fig:tcanther}
\end{figure}

The replacement channel (Fig.~\ref{fig:tcanther}) can be implemented using its Kraus decomposition: \[\mathcal{E}_{\sigma}(\rho) = \sum_{i,j}K_{i,j}\rho K_{i,j}^{\dagger}\] where $K_{i,j}$ are Kraus operators are defined as $$K_{i,j} = \sqrt{\lambda_i}\ket{\psi_i}\bra{j}$$ with $\sigma = \sum_i \lambda_i\ket{\psi_i}\bra{\psi_i}$ being an eigendecomposition of a target density matrix and $ \sum_j \ket{j}\bra{j} = I $ being is an orthonormal basis for the Hilbert Space.

Using this definition of the Kraus operators, it is trivial to check that the Kraus operators satisfy the condition $  \sum_{i,j}K_{i,j}^\dagger K_{i,j} = I$
and $\mathcal{E}_{\sigma}(\rho) = \sum_{i,j}K_{i,j}\rho K_{i,j}^{\dagger}= \sum_{ij} \lambda_i  \ket{\psi_i}\bra{j} \rho \ket{j}\bra{\psi_i} = (\sum_i \lambda_i \ket{\psi_i} \bra{\psi_i}) (\sum_j \bra{j} \rho \ket{j}) = Tr[\rho] \sigma$

\section*{Acknowledgment}
N.S. was supported by funding from the National Science Foundation award DMS 2015431 (for INMAS Midwest). This material is based upon work supported by the U.S. Department of Energy, Office Science, Advanced Scientific Computing Research (ASCR) program under contract number DE-AC02-06CH11357 as part of the InterQnet quantum networking project. We acknowledge Sean E. Sullivan for helpful discussions on remote gate simulation and valuable feedback on the manuscript.

\noindent\framebox{\parbox{0.97\linewidth}{
The submitted manuscript has been created by UChicago Argonne, LLC, Operator of 
Argonne National Laboratory (``Argonne''). Argonne, a U.S.\ Department of 
Energy Office of Science laboratory, is operated under Contract No.\ 
DE-AC02-06CH11357. 
The U.S.\ Government retains for itself, and others acting on its behalf, a 
paid-up nonexclusive, irrevocable worldwide license in said article to 
reproduce, prepare derivative works, distribute copies to the public, and 
perform publicly and display publicly, by or on behalf of the Government.  The 
Department of Energy will provide public access to these results of federally 
sponsored research in accordance with the DOE Public Access Plan. 
http://energy.gov/downloads/doe-public-access-plan.}}

\nocite{}
\bibliographystyle{unsrt}
\bibliography{Bibliography}% Produces the bibliography via BibTeX.

\end{document}